\documentclass[letter]{aa}
\usepackage{graphicx}
\usepackage{txfonts}
\begin{document}
\title{
Inhomogeneities in molecular layers of Mira atmospheres
\thanks{Based on observations made with the VLT Interferometer (VLTI)
at Paranal Observatory under program ID 082.D-0723}}
\author{
M.~Wittkowski\inst{1} \and
D.~A.~Boboltz\inst{2} \and
M.~Ireland\inst{3,4} \and
I.~Karovicova\inst{1} \and
K.~Ohnaka\inst{5} \and
M.~Scholz\inst{6,7}\and
F.~van~Wyk\inst{8}\and
P.~Whitelock\inst{8,9} \and
P.~R.~Wood\inst{10} \and
A.~A.~Zijlstra\inst{11}
}
\institute{
ESO, Karl-Schwarzschild-Str. 2,
85748 Garching bei M\"unchen, Germany,
\email{mwittkow@eso.org}
\and
US Naval Observatory, 3450 Massachusetts Avenue,
NW, Washington, DC 20392-5420, USA
\and
Department of Physics and Astronomy, Macquarie University  NSW  2109,
Australia
\and
Australian Astronomical Observatory, PO Box 296, Epping, NSW 1710, Australia
\and
Max-Planck-Institut f\"ur Radioastronomie,
Auf dem H\"ugel 69, 53121 Bonn, Germany
\and
Zentrum f\"ur Astronomie der Universit\"at Heidelberg (ZAH),
Institut f\"ur Theoretische Astrophysik, 
Albert-Ueberle-Str. 2, 69120 Heidelberg, Germany 
\and
Sydney Institute for Astronomy, School of Physics,
University of Sydney, Sydney NSW 2006, Australia
\and
South African Astronomical Observatory (SAAO), PO Box 9, 
7935 Observatory, South Africa
\and
Astrophysics, Cosmology and Gravitation Centre, Astronomy Dept, 
University of Cape Town, 7701 Rondebosch, South Africa
\and
Research School of Astronomy and Astrophysics, 
The Australian National University, Canberra, Australia
\and
School of Physics \& Astronomy, University of Manchester,
P.O. Box 88, Manchester M60 1QD, UK
}
\date{Received \dots; accepted \dots}
\abstract{}
{We investigate the  structure and shape of the photospheric 
and molecular layers of the atmospheres of four Mira variables.}
{We obtained near-infrared $K$-band spectro-interferometric observations
of the Mira variables \object{R~Cnc}, \object{X~Hya}, 
\object{W~Vel}, and \object{RW~Vel} with a spectral 
resolution of about 1500 using the AMBER instrument at the VLTI. We obtained
concurrent JHKL photometry using the the Mk~II instrument at the SAAO.
}
{The Mira stars in our sample are found to have wavelength-dependent visibility values 
that are consistent with earlier low-resolution AMBER observations of S Ori 
and with the predictions of dynamic model atmosphere series based on 
self-excited pulsation models. The corresponding wavelength-dependent 
uniform disk (UD) diameters show a minimum near the near-continuum bandpass
at 2.25\,$\mu$m. They then increase by up to 30\% toward the H$_2$O band
at 2.0\,$\mu$m and by up to 70\% at the CO bandheads between
2.29\,$\mu$m and 2.48\,$\mu$m.
The dynamic model atmosphere series show a consistent wavelength-dependence, 
and their parameters such as the visual phase, effective temperature,
and distances are consistent with independent estimates.
The closure phases have significantly wavelength-dependent 
and non-zero values at all wavelengths indicating deviations from 
point symmetry. For example, the R Cnc closure phase
is 110\degr\ $\pm$ 4\degr\ in the 2.0\,$\mu$m H$_2$O band, 
corresponding for instance to an additional unresolved 
spot contributing 3\% of the total flux at 
a separation of $\sim$ 4\,mas.
}
{Our observations are consistent with the predictions of the latest
dynamic model atmosphere series based on self-excited pulsation models. The 
wavelength-dependent radius variations are interpreted as the effect of 
molecular layers lying above the photosphere. The wavelength-dependent 
closure phase values are indicative of deviations from point symmetry at all wavelengths,
thus a complex non-spherical stratification of the extended atmosphere.
In particular, the significant deviation from point symmetry in the H$_2$O 
band is interpreted as a signature on large scales (there being a few across 
the stellar disk) of inhomogeneities or clumps in 
the water vapor layer. The observed inhomogeneities might possibly be caused 
by pulsation- and shock-induced chaotic motion in the extended atmosphere.
}
\maketitle
\section{Introduction}
Mira stars are long-period, large-amplitude variable stars on the asymptotic 
giant branch (AGB). Mass loss becomes increasingly important toward the tip 
of the AGB, before the star evolves to the planetary nebula (PN) phase, where
a great diversity of morphologies is seen. Contemporary astrophysical 
questions related to the study of Mira variables include the pursuit of the 
detailed mass-loss mechanism on the AGB, including the effects of pulsation 
and shock fronts on the structure and morphology of the extended atmosphere, 
and of the mechanism shaping the observed PN morphologies.
Near-infrared interferometry has traditionally been used to characterize 
AGB star atmospheres. In particular, observations using the IOTA interferometer
have uncovered the wavelength-dependence of Mira star diameters using a few
bandpasses with spectral resolutions of up to $\lambda/\Delta\lambda\sim25$, 
providing observational evidence of molecular layers lying outside the 
photospheric layers (e.g.; Mennesson et al.  \cite{mennesson02}; 
Perrin et al. \cite{perrin04}). These molecular layers have also been present 
in theoretical dynamic model atmospheres (Hofmann et al. \cite{hofmann98}, 
Ireland et al. \cite{ireland04a,ireland04b}). $H$-band interferometry at the 
IOTA interferometer in the broad band (Ragland et al. \cite{ragland06}), as 
well as in three filters with $\lambda/\Delta\lambda\sim15$ (Ragland et 
al. \cite{ragland08}, Pluzhnik et al. \cite{pluzhnik09}) have revealed 
non-zero closure phases for several Mira variables, which reflect asymmetric 
brightness distributions of the photosphere and/or the envelope around the 
star. Wittkowski et al. (\cite{wittkowski08}) presented the first VLTI/AMBER 
near-infrared spectro-interferometric observation of an AGB star, providing 
continuous wavelength coverage from 1.29\,$\mu$m to 2.32\,$\mu$m with a 
spectral resolution of $\lambda/\Delta\lambda\sim35$. The data showed 
visibility and diameter variations as a function of wavelength that generally 
confirmed the predictions of dynamic model atmospheres, where the diameter 
variations can be understood as the effects from atmospheric molecular layers 
(most importantly H$_2$O, CO, and TiO).
Here, we present the first VLTI/AMBER observations of Mira variables using 
its $K$ medium resolution modes with a spectral resolution of 
$\lambda/\Delta\lambda\sim1500$, and a comparison to the newly available 
{\tt CODEX} dynamic model-atmosphere series by Ireland et al. (\cite{ireland08,ireland11}),
which are based on self-excited pulsation models.
\section{Observations} 
\begin{table}
\caption{Log of our VLTI/AMBER observations}
\label{tab:vlti}
\centering
\begin{tabular}{lrrrrrrrrr}
\hline\hline
Target &  Date       &  JD        & Mode    & Baseline        & PA     \\ 
       &             &            &         & m               & $\deg$ \\\hline
R Cnc  & 2008-12-29  &  4830      & MR23    & 15.9/31.8/47.7  &-74     \\
       & 2008-12-30  &  4831      & MR21    & 16.0/32.0/48.0  &-73     \\
X Hya  & 2008-12-29  &  4830      & MR23    & 16.0/31.9/47.9  &-71     \\
       & 2008-12-30  &  4831      & MR21    & 15.5/30.9/46.4  &-76     \\
W Vel  & 2008-12-29  &  4830      & MR23    & 15.3/30.6/45.9  &-84     \\
RW Vel & 2009-03-01  &  4892      & MR23    & 15.1/30.1/45.2  &-86     \\\hline
\end{tabular}
\tablefoot{
The Julian date (JD) is given as JD-2450000.
The AMBER instrument mode is MR23 (medium resolution mode with
central wavelength 2.3\,$\mu$m and range 2.12--2.47\,$\mu$m) or 
MR21 (2.1\,$\mu$m, 1.92--2.26\,$\mu$m).
The baseline is the projected baseline length for the AT VLTI baselines used
(E0-G0/G0-H0/E0-H0) with projected baseline angle PA.
}
\end{table}
\begin{table}
\caption{SAAO JHKL photometry}
\label{tab:saao}
\centering
\begin{tabular}{lrrrrrrrr}
\hline\hline
Target & $J$   & $H$    & $K$    & $L$     & $A_V$ & $m_\mathrm{bol}$ \\
       & mag   &  mag   &  mag   &  mag    & mag   & mag              \\\hline
 R Cnc & 0.866 & -0.155 & -0.652 & -1.063  & 0.07  & 2.55             \\
 X Hya & 2.375 &  1.476 &  0.982 &  0.397  & 0.18  & 4.12             \\
 W Vel & 2.002 &  0.987 &  0.503 & -0.097  & 0.42  & 3.64             \\\hline
\end{tabular}
\tablefoot{The observations were obtained on 2008 December 30 (JD 2454831),
concurrently with the AMBER observations in Table~\ref{tab:vlti}.
}
\end{table}
We obtained $K$-band spectro-interferometry with a resolution of 
$\lambda/\Delta\lambda\sim1500$ of the Mira variables R~Cnc, X~Hya, W~Vel, 
and RW~Vel with the AMBER instrument (Petrov et al. \cite{petrov07}) of the  
VLTI. Table~\ref{tab:vlti} shows the details of our observations. We obtained 
concurrent near-infrared $JHKL$ photometry of R~Cnc, X~Hya, and W~Vel at the 
SAAO Mk~II instrument. Table~\ref{tab:saao} lists the obtained photometry and 
the estimated bolometric magnitudes using the procedures outlined in 
Whitelock et al. (\cite{whitelock08}) and their estimates of $A_V$. For R~Cnc 
and X~Hya, we obtained additional AMBER and SAAO observations at later 
epochs/phases, as well as coordinated mid-infrared VLTI/MIDI observations and 
VLBA observations of the SiO and H$_2$O maser emission. The complete data 
set for these sources will be discussed in a dedicated forthcoming paper.

Raw visibility and closure phase values were obtained from the AMBER data 
using version 3.0 of the {\tt amdlib} data reduction package (Tatulli et 
al. \cite{tatulli07}, Chelli et al. \cite{chelli09}). An absolute wavelength 
calibration was performed by correlating the AMBER flux spectra with a 
reference spectrum that includes the AMBER transmission curve, the telluric 
spectrum estimated with ATRAN (Lord \cite{lord92}), and the expected stellar 
spectrum using the BS~4432 spectrum from Lan{\c c}on \& Wood (\cite{lancon00}),
which has a similar spectral type as our calibrators (K3-4 III). Similarly, a 
relative flux calibration of the target stars was performed using the 
calibrator stars and the BS~4432 spectrum. Calibrated visibility spectra were
obtained by using an average of two transfer function measurements taken 
before and after each science target observation.  The calibrators were 
$\iota$~Hya (for R Cnc and X~Hya), 81~Gem (for R~Cnc), 31~Leo (for X~Hya), 
GZ~Vel (for W~Vel and RW~Vel), and $\gamma$~Pyx (for W Vel), and were 
selected from Bord{\'e} et al. (\cite{borde02}) using the ESO calibrator 
tool CalVin.
\section{Results and comparison to model atmospheres}
\begin{figure}
\centering
\resizebox{1\hsize}{!}{\includegraphics{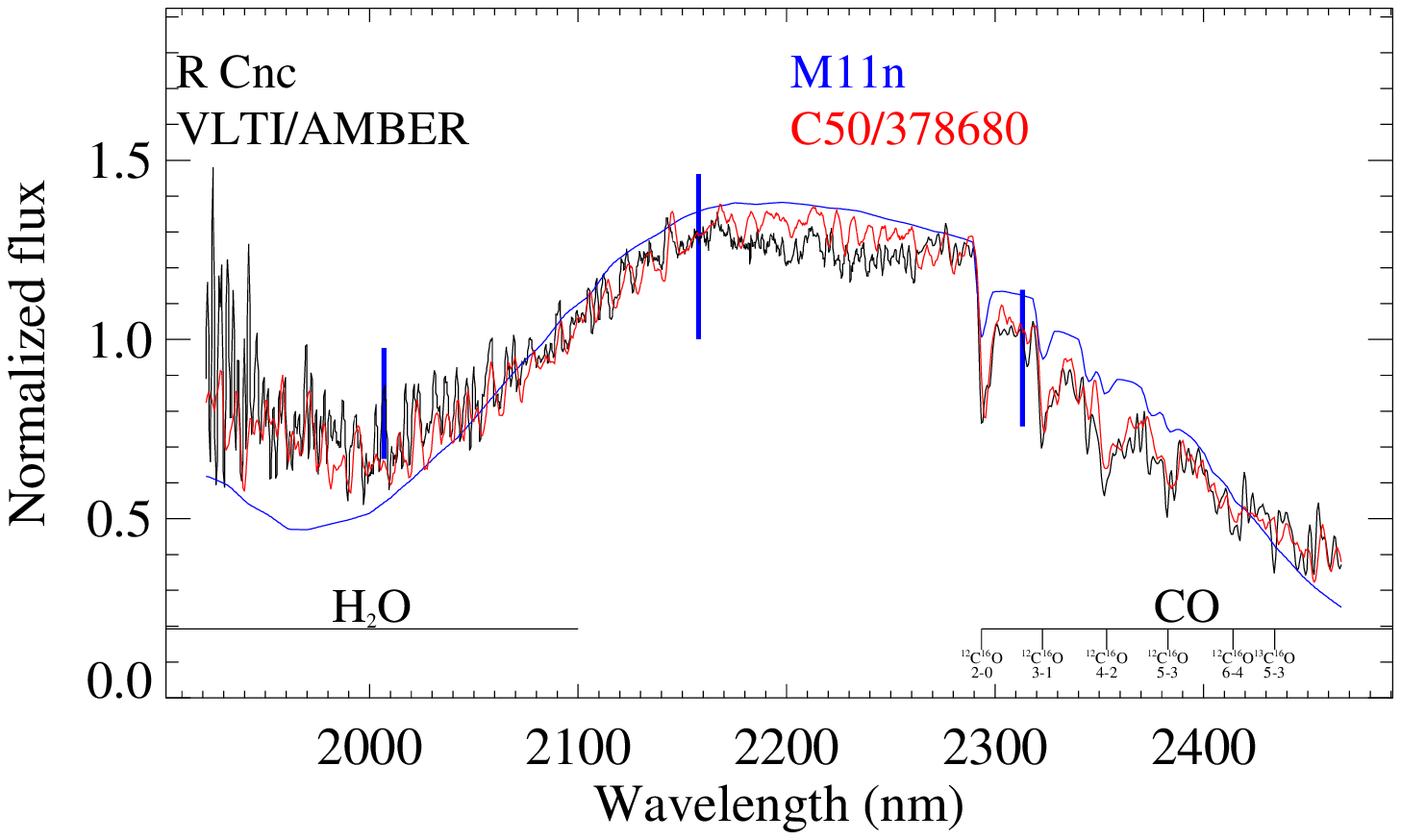}}

\resizebox{1\hsize}{!}{\includegraphics{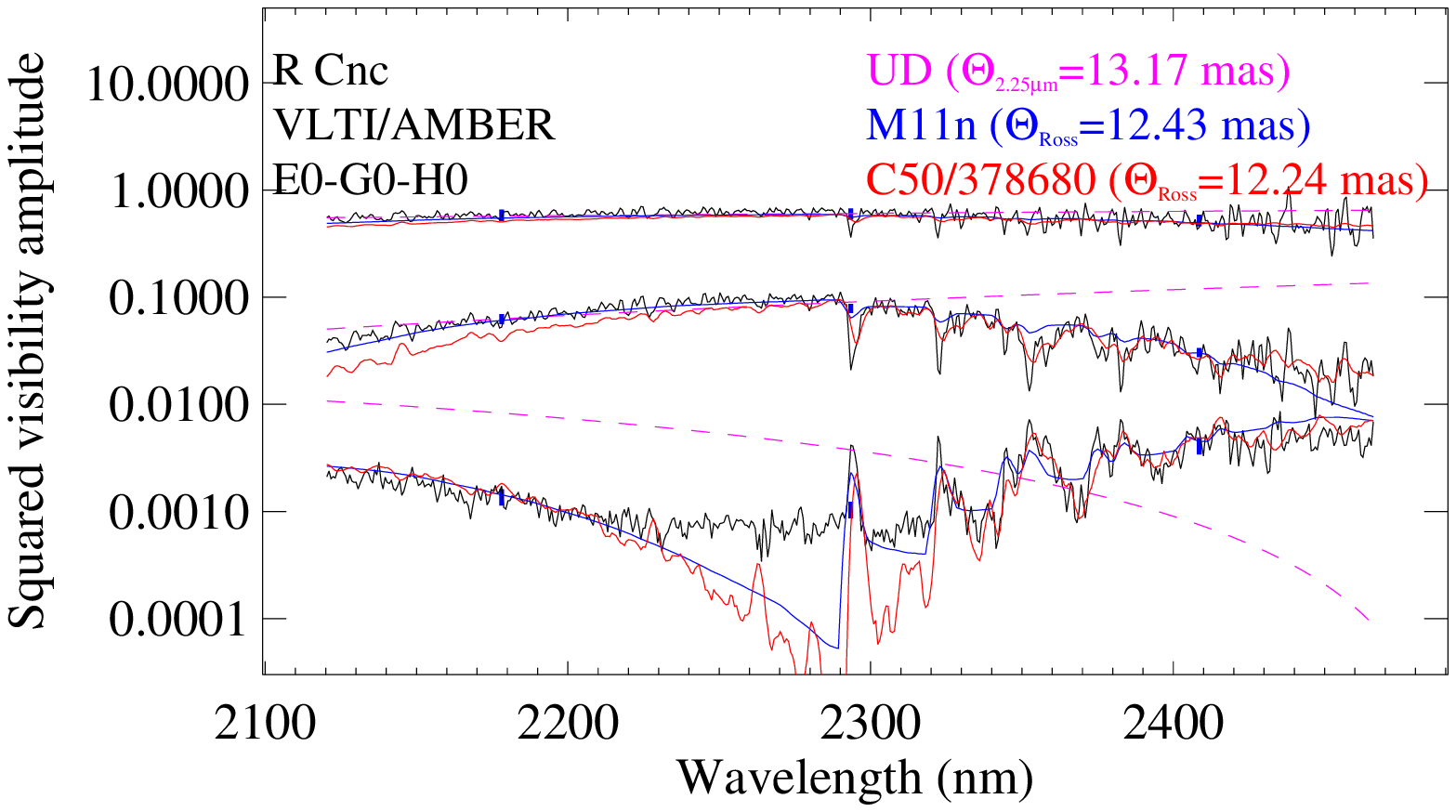}}

\resizebox{1\hsize}{!}{\includegraphics{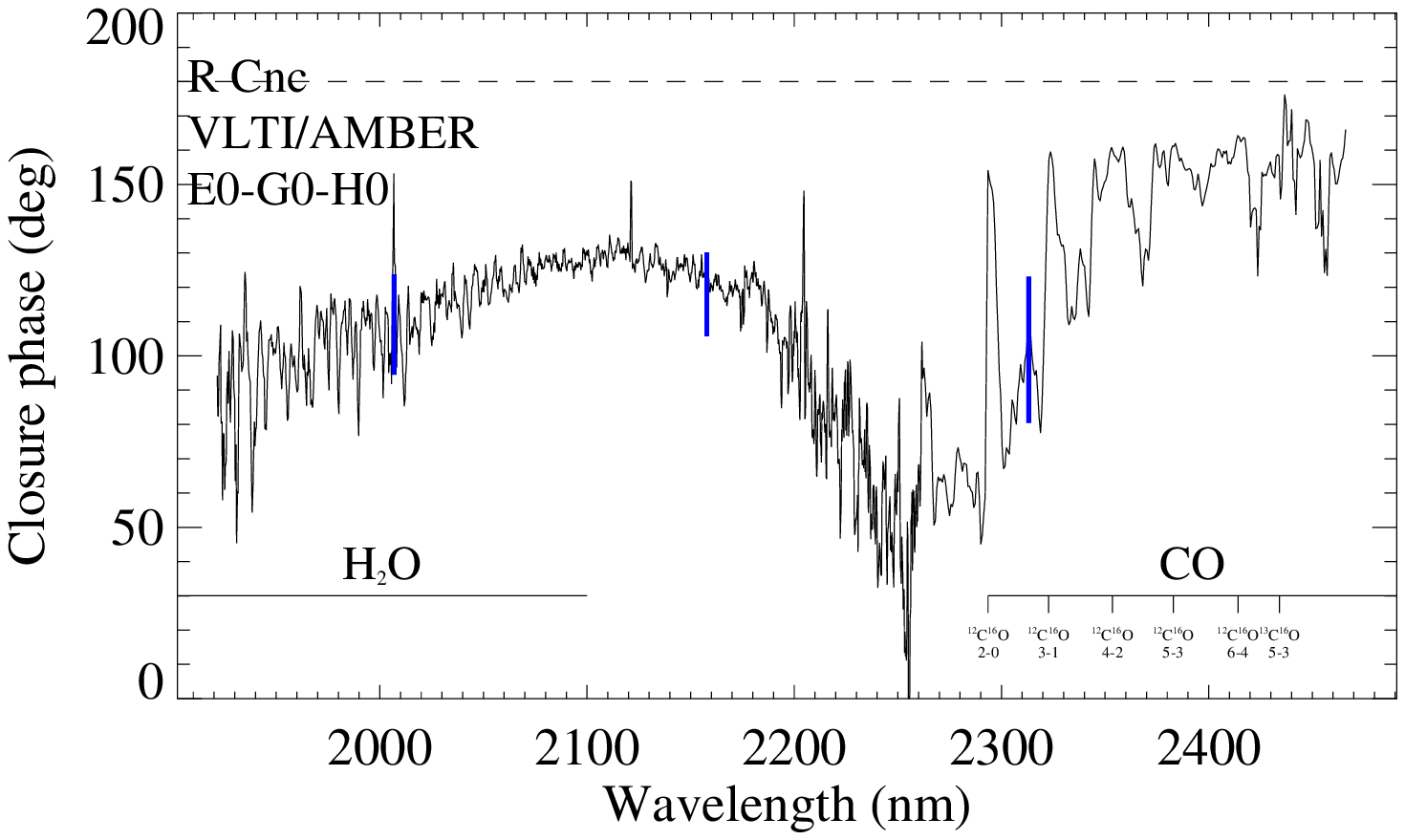}}

\resizebox{1\hsize}{!}{\includegraphics{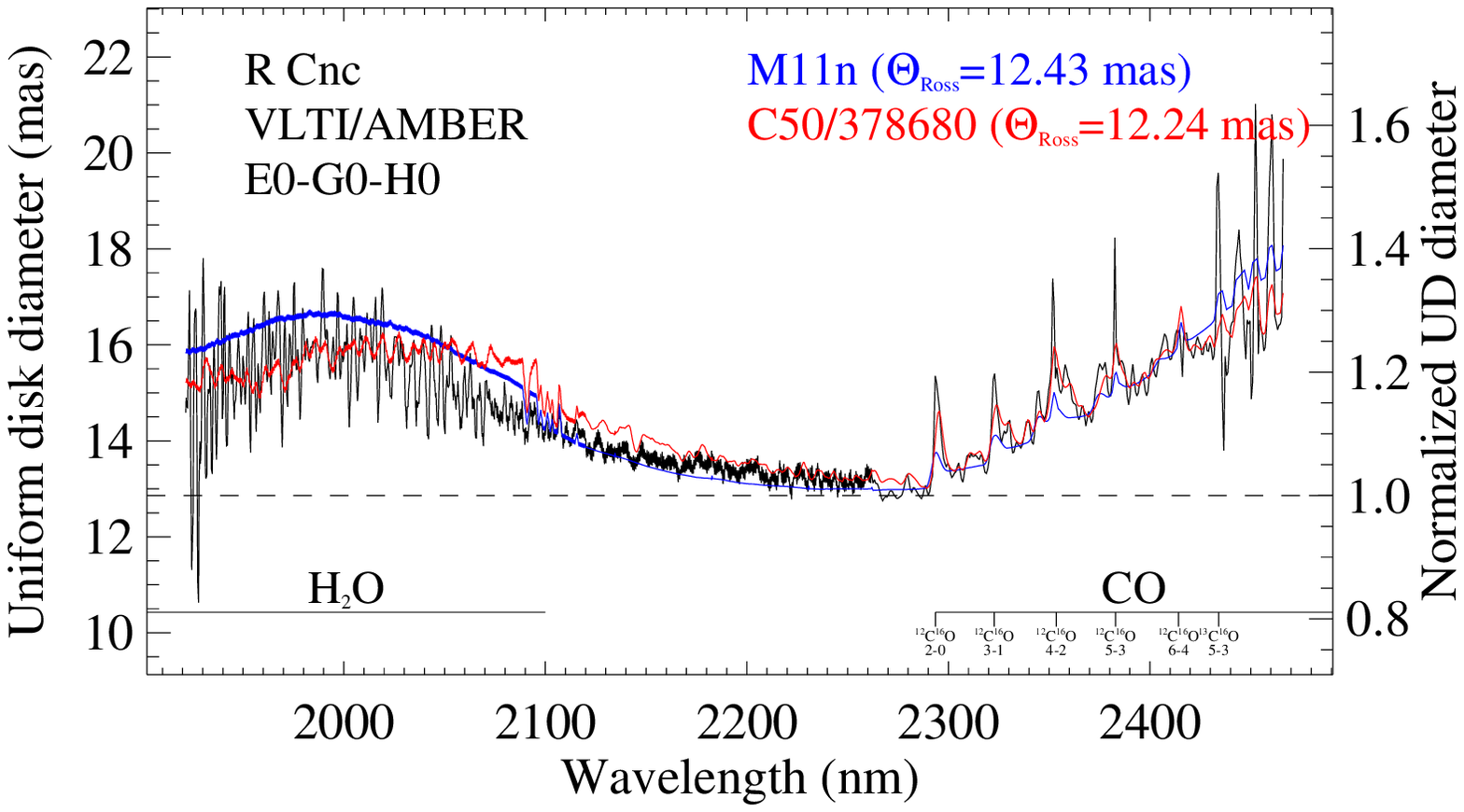}}
\caption{R~Cnc flux, squared visibility amplitude, 
closure phase, and UD diameter (from top to bottom)
as a function of wavelength.
Data of the remaining three targets are shown in Fig. 2 of the 
electronic version.
The wavelength range
is 1.92--2.47\,$\mu$m (MR21 mode: 1.92--2.26\,$\mu$m,
MR23 mode: range 2.12--2.47\,$\mu$m). The modes overlap
in the region 2.12--2.26\,$\mu$m, where the two lines
appear as one thicker line.
The thick vertical
lines denote the errors averaged over three wavelength intervals.
The blue lines show a comparison to the best fitting model of the
{\tt P/M} model atmosphere series 
(Ireland et al. \cite{ireland04a,ireland04b}), 
and the red lines the best fitting model of
the new {\tt CODEX} series (Ireland et al. \cite{ireland08,ireland11}). 
The dashed line shows a uniform 
disk curve 
fitted to the 2.25\,$\mu$m near-continuum bandpass.}
\label{fig:miras}
\end{figure}

\onlfig{2}{
\begin{figure*}
\centering
\resizebox{0.32\hsize}{!}{\includegraphics{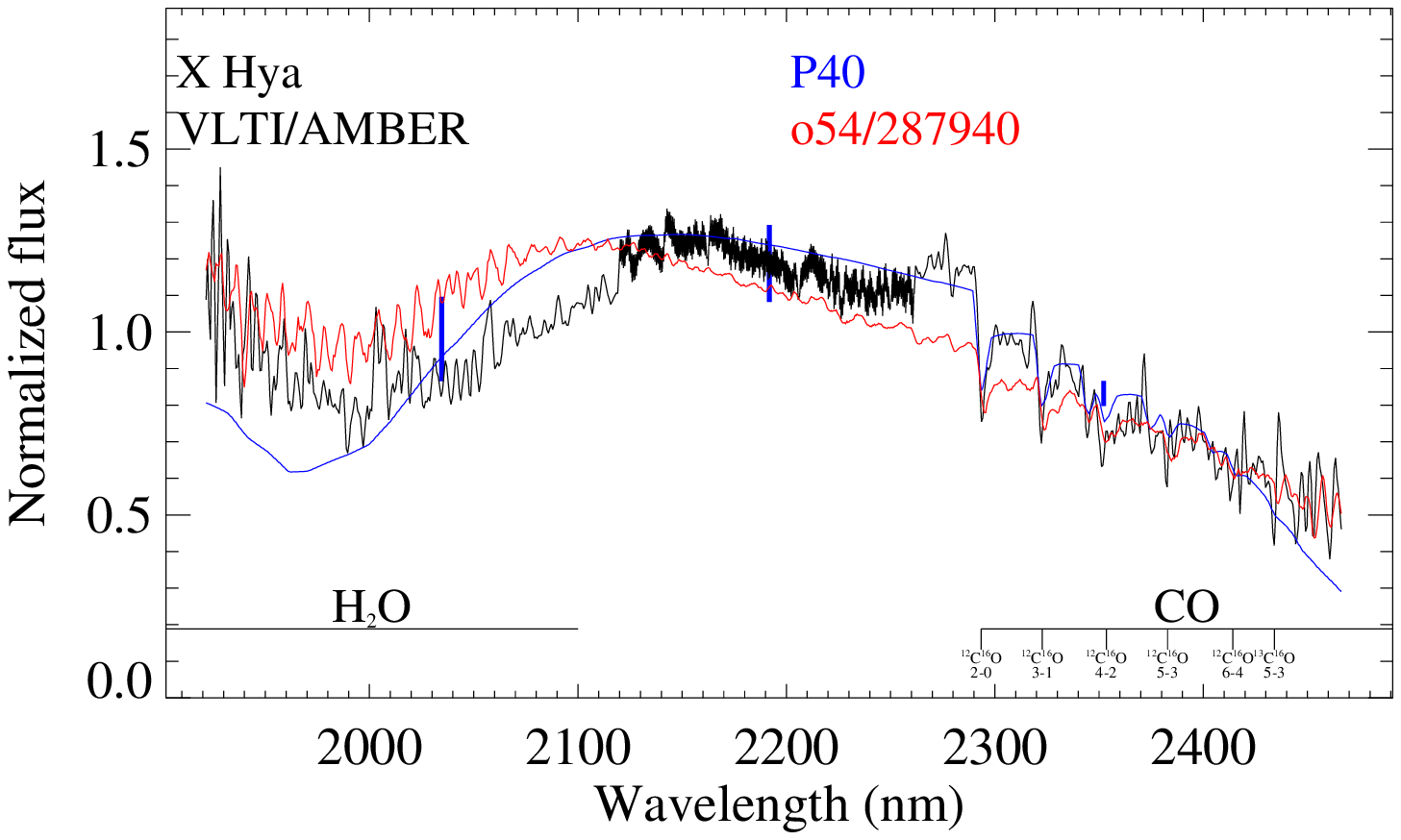}}
\resizebox{0.32\hsize}{!}{\includegraphics{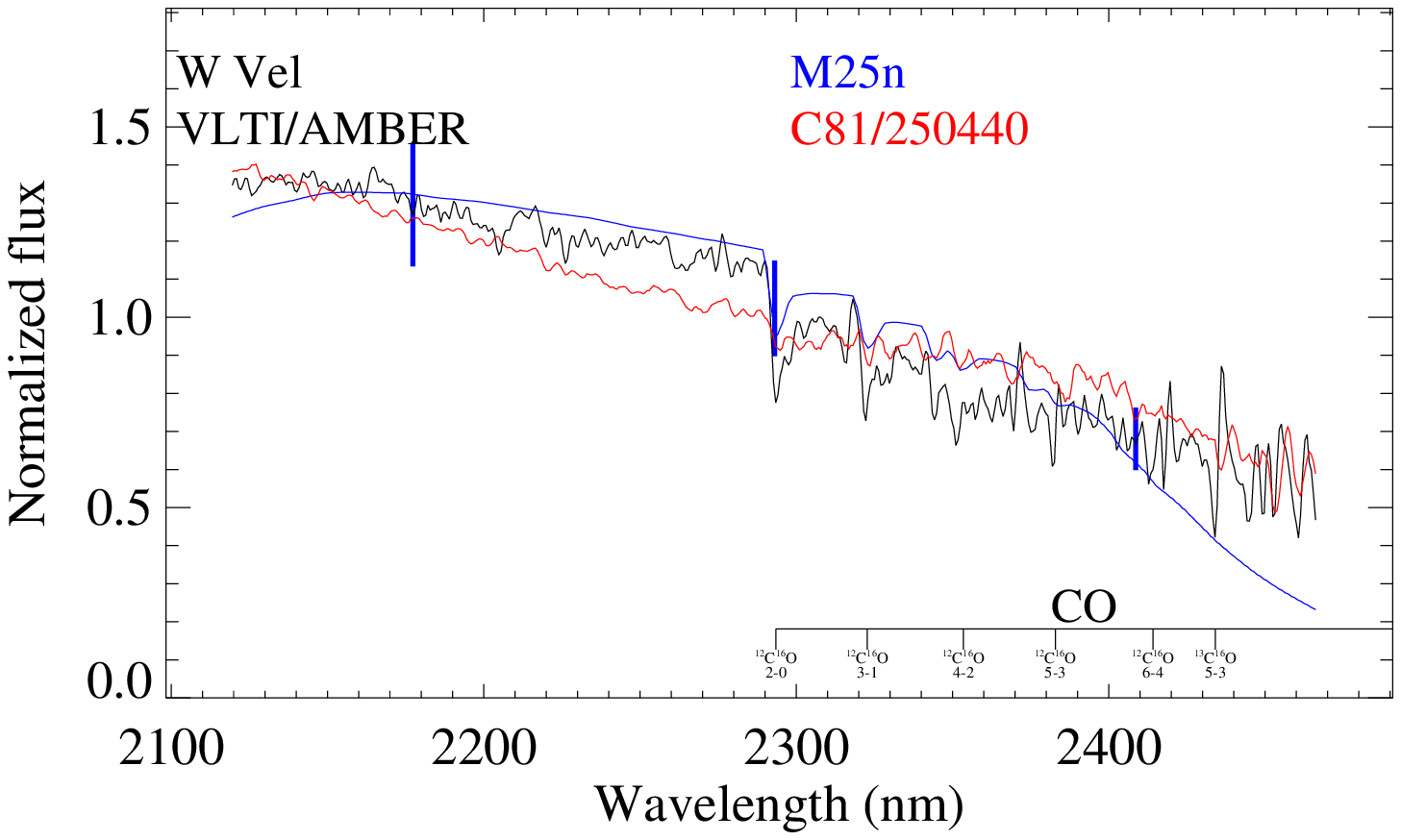}}
\resizebox{0.32\hsize}{!}{\includegraphics{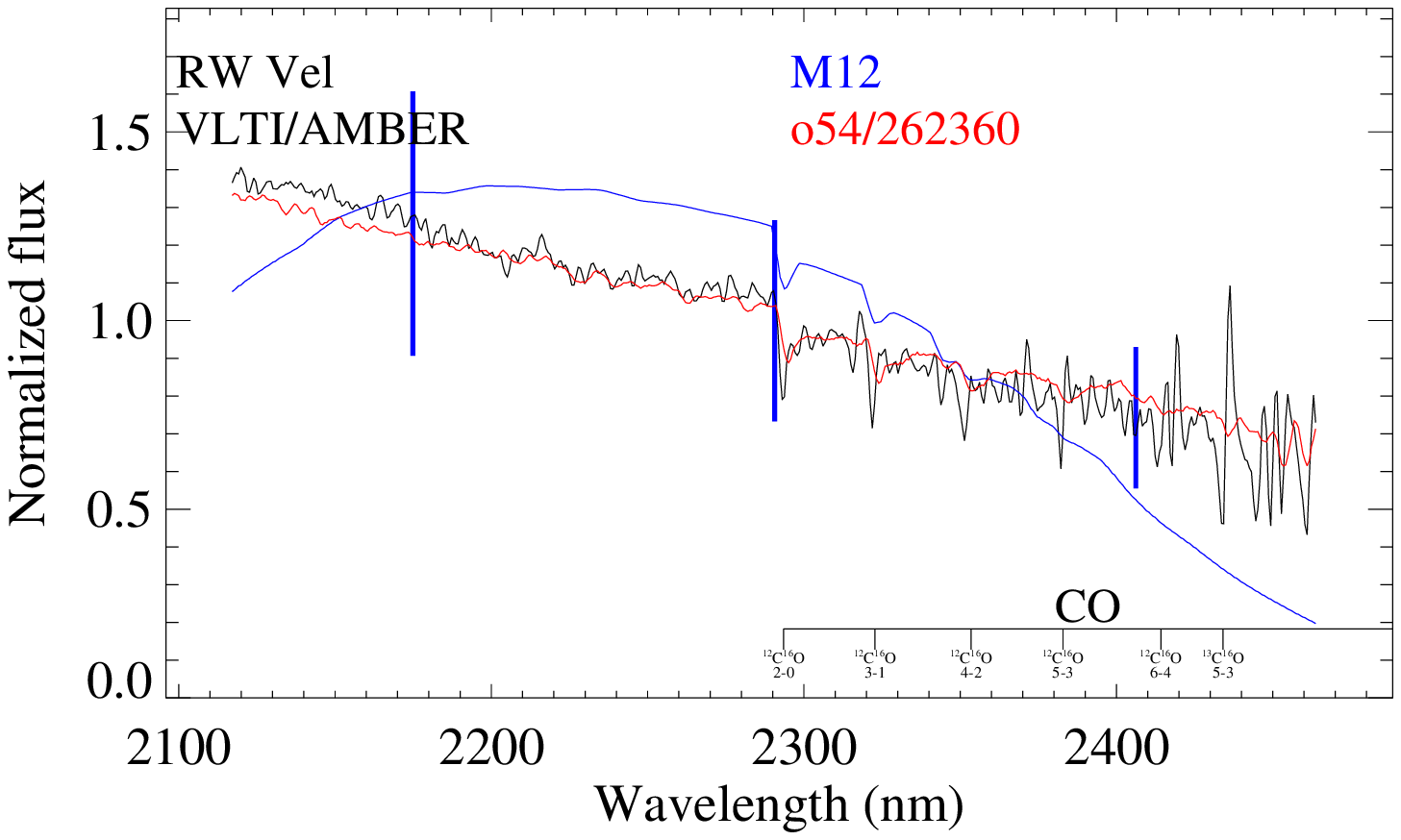}}

\resizebox{0.32\hsize}{!}{\includegraphics{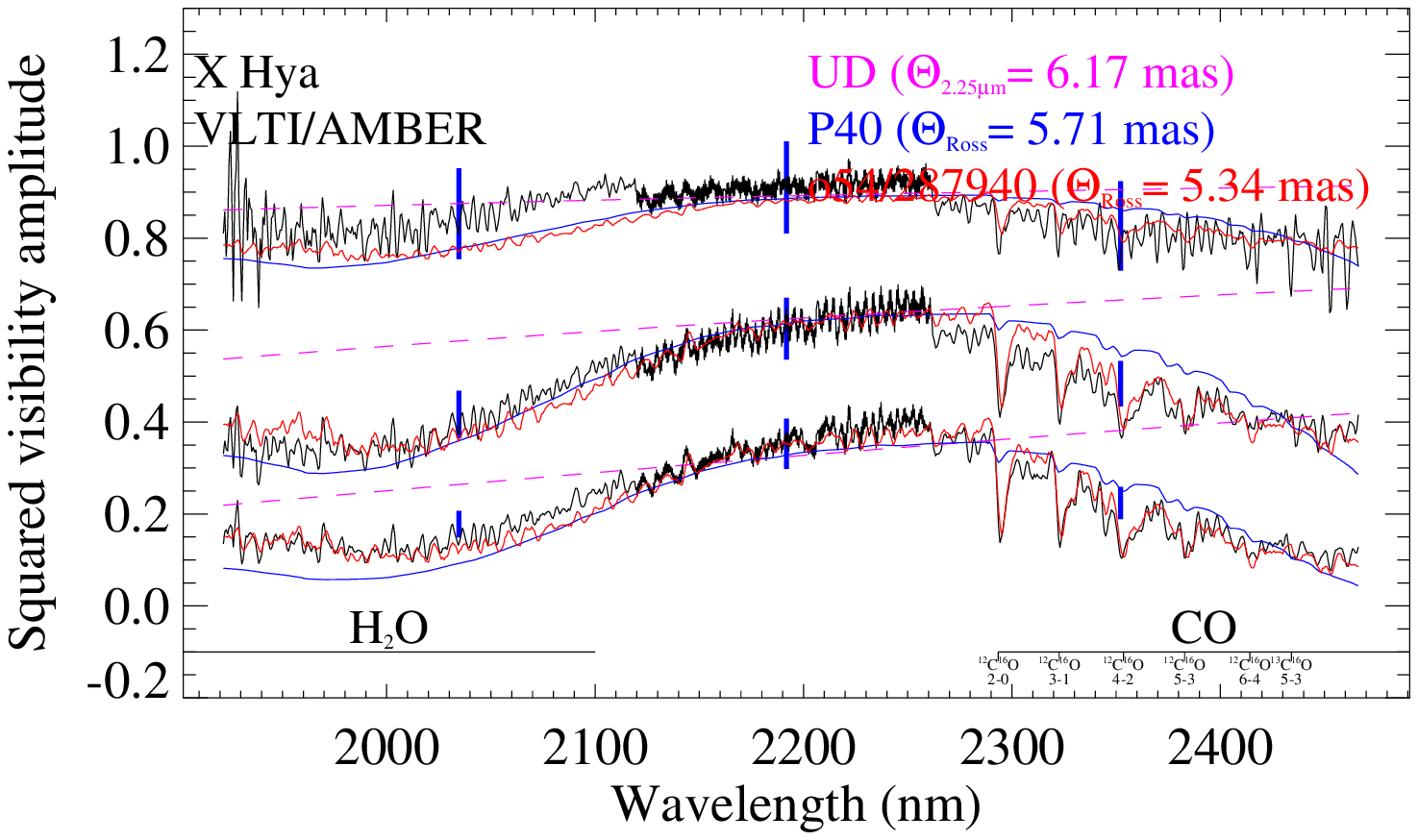}}
\resizebox{0.32\hsize}{!}{\includegraphics{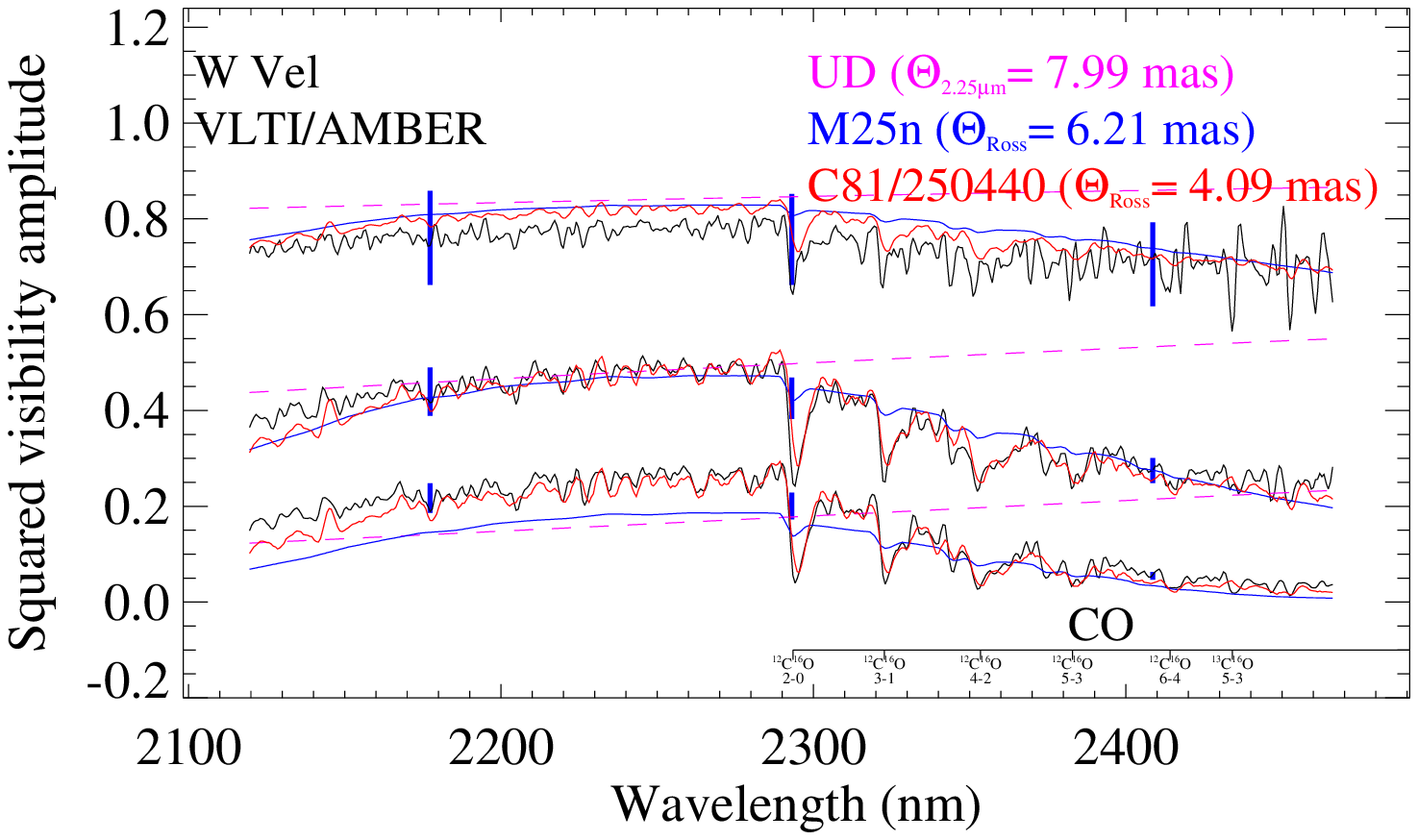}}
\resizebox{0.32\hsize}{!}{\includegraphics{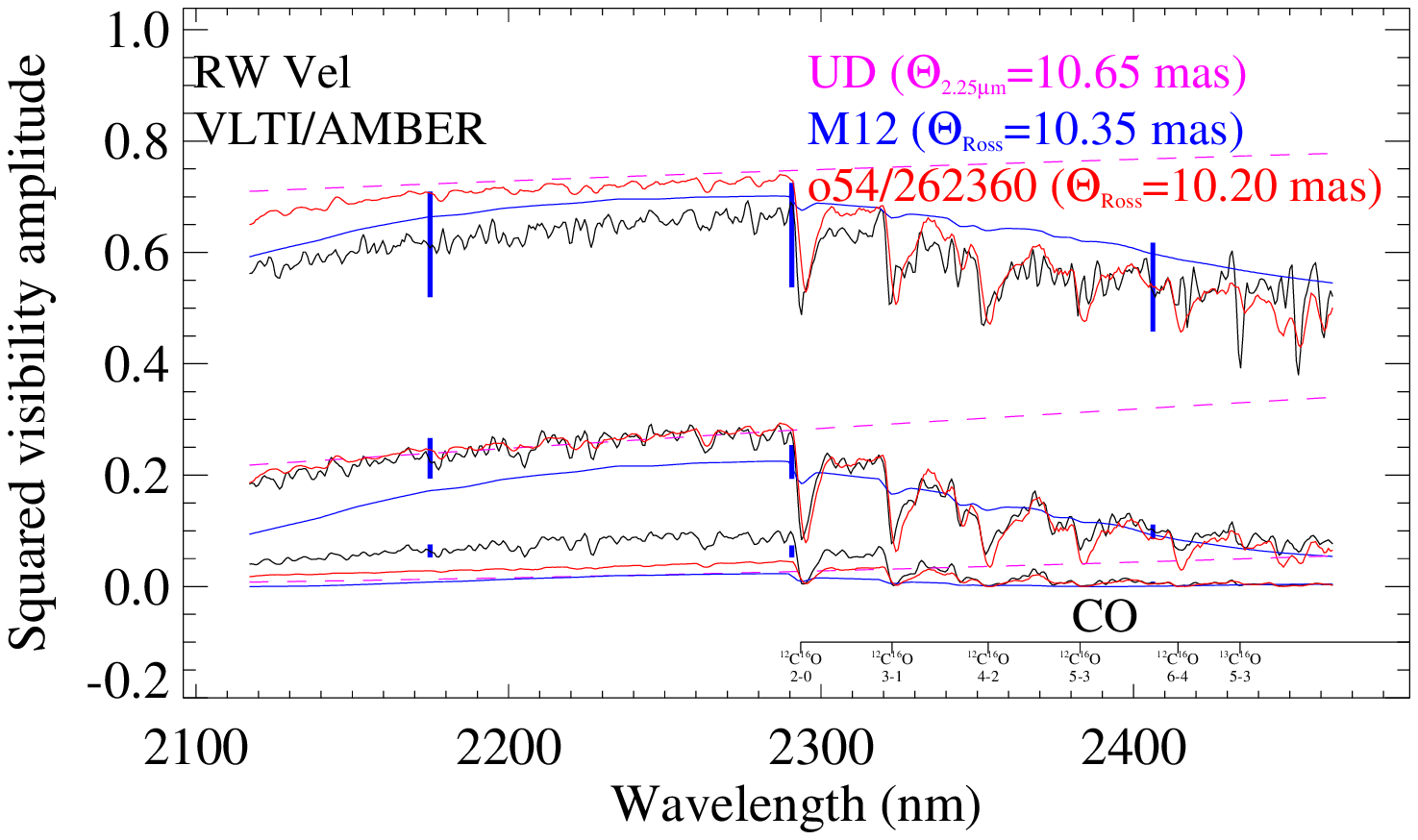}}

\resizebox{0.32\hsize}{!}{\includegraphics{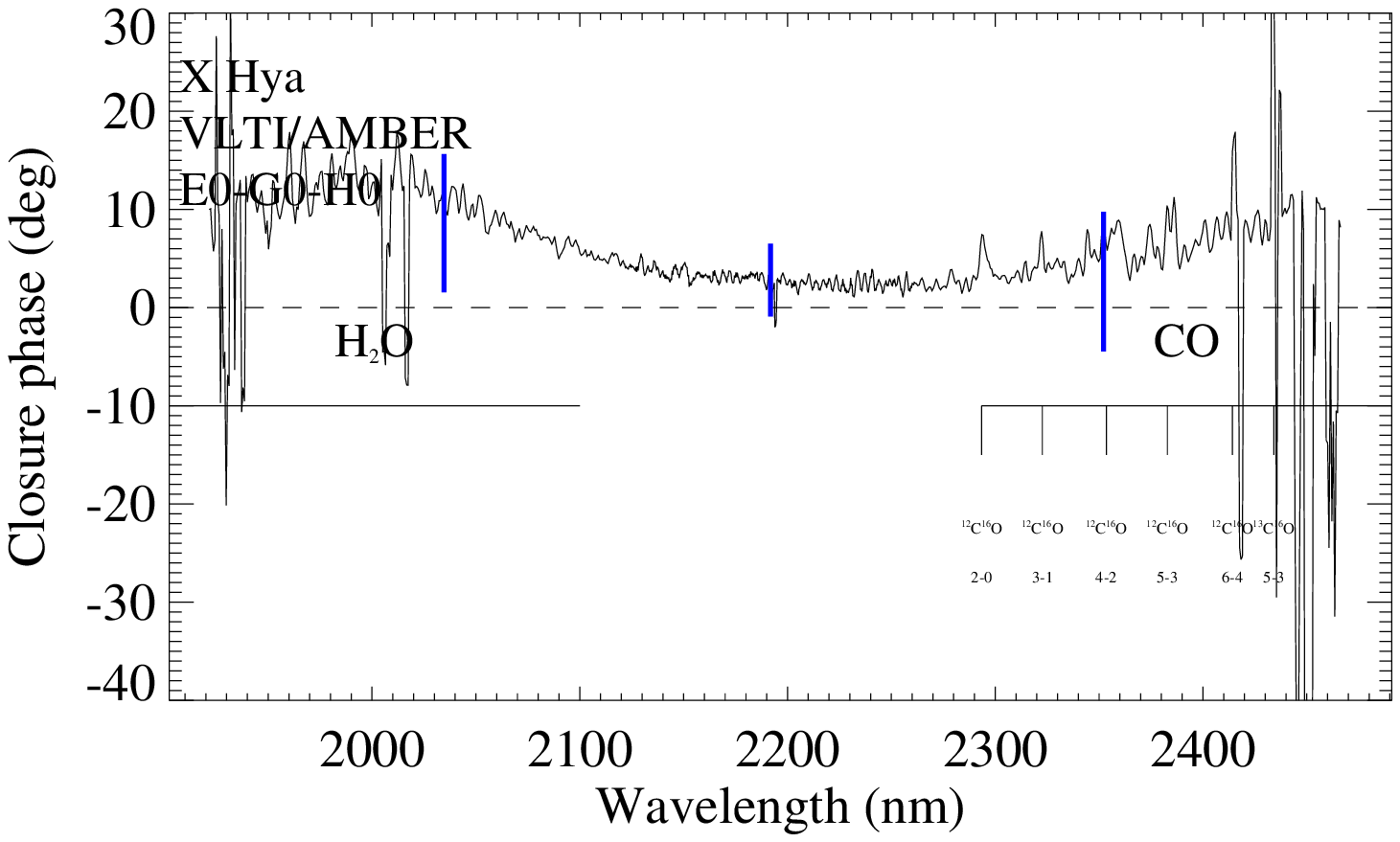}}
\resizebox{0.32\hsize}{!}{\includegraphics{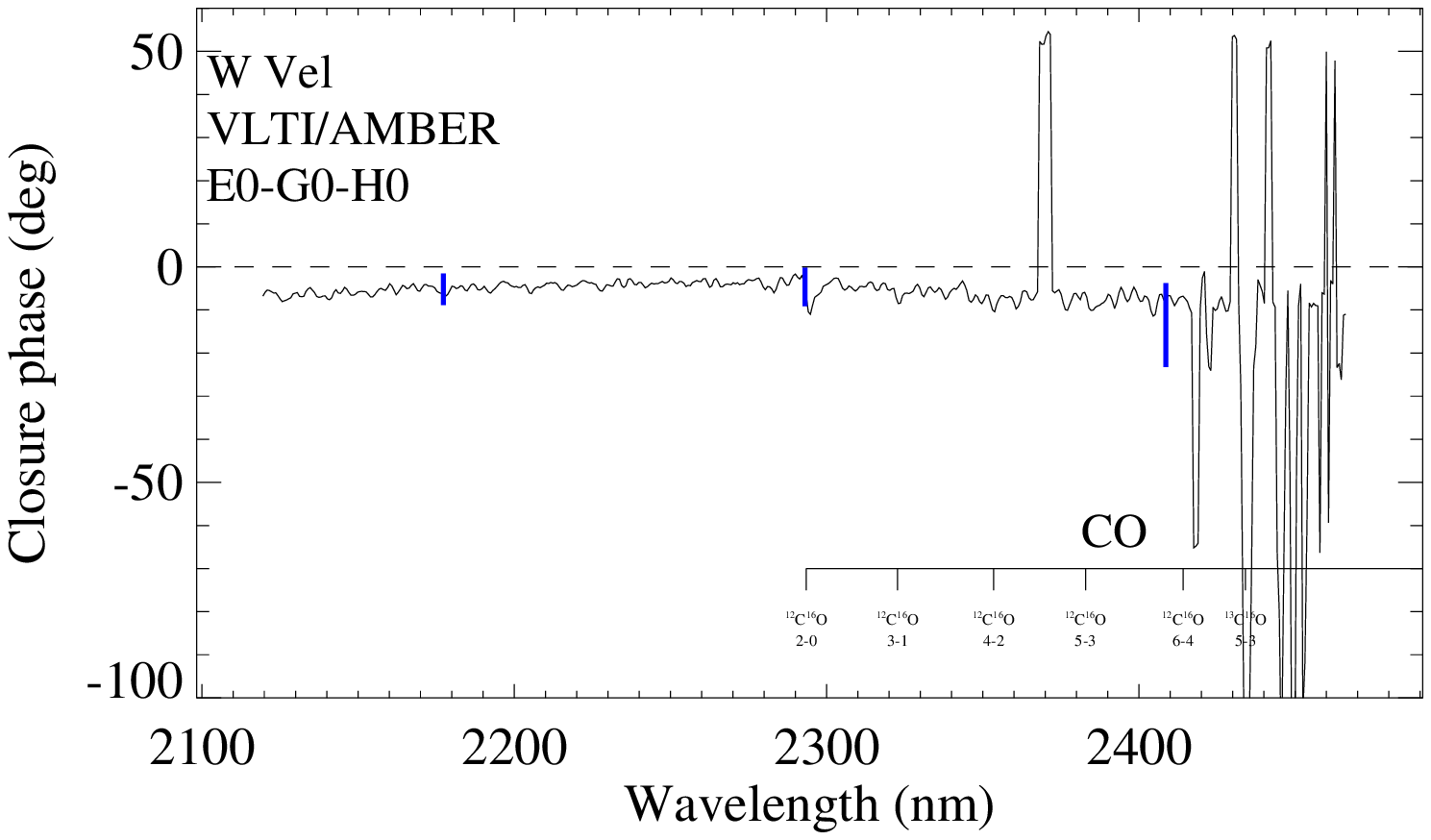}}
\resizebox{0.32\hsize}{!}{\includegraphics{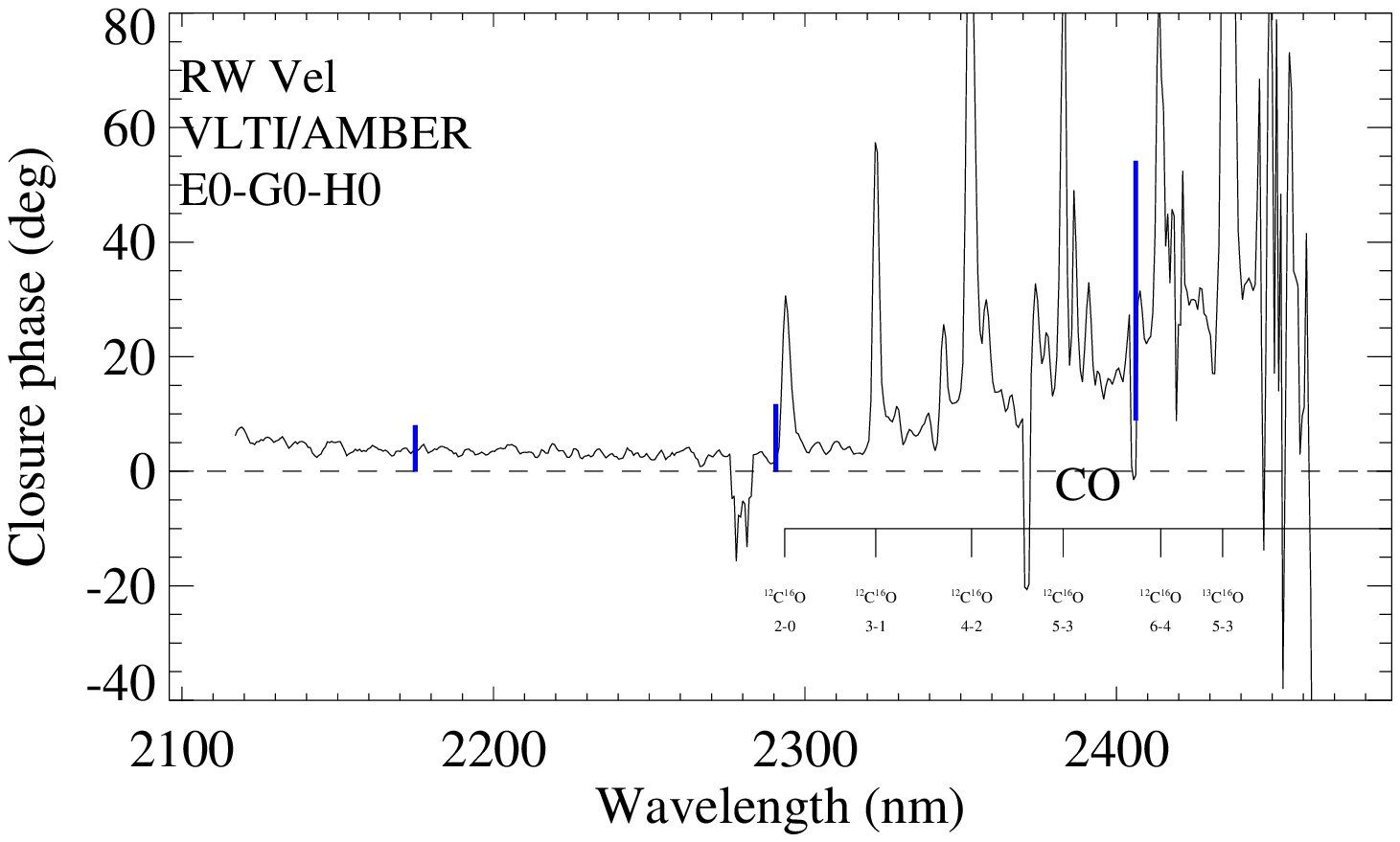}}

\resizebox{0.32\hsize}{!}{\includegraphics{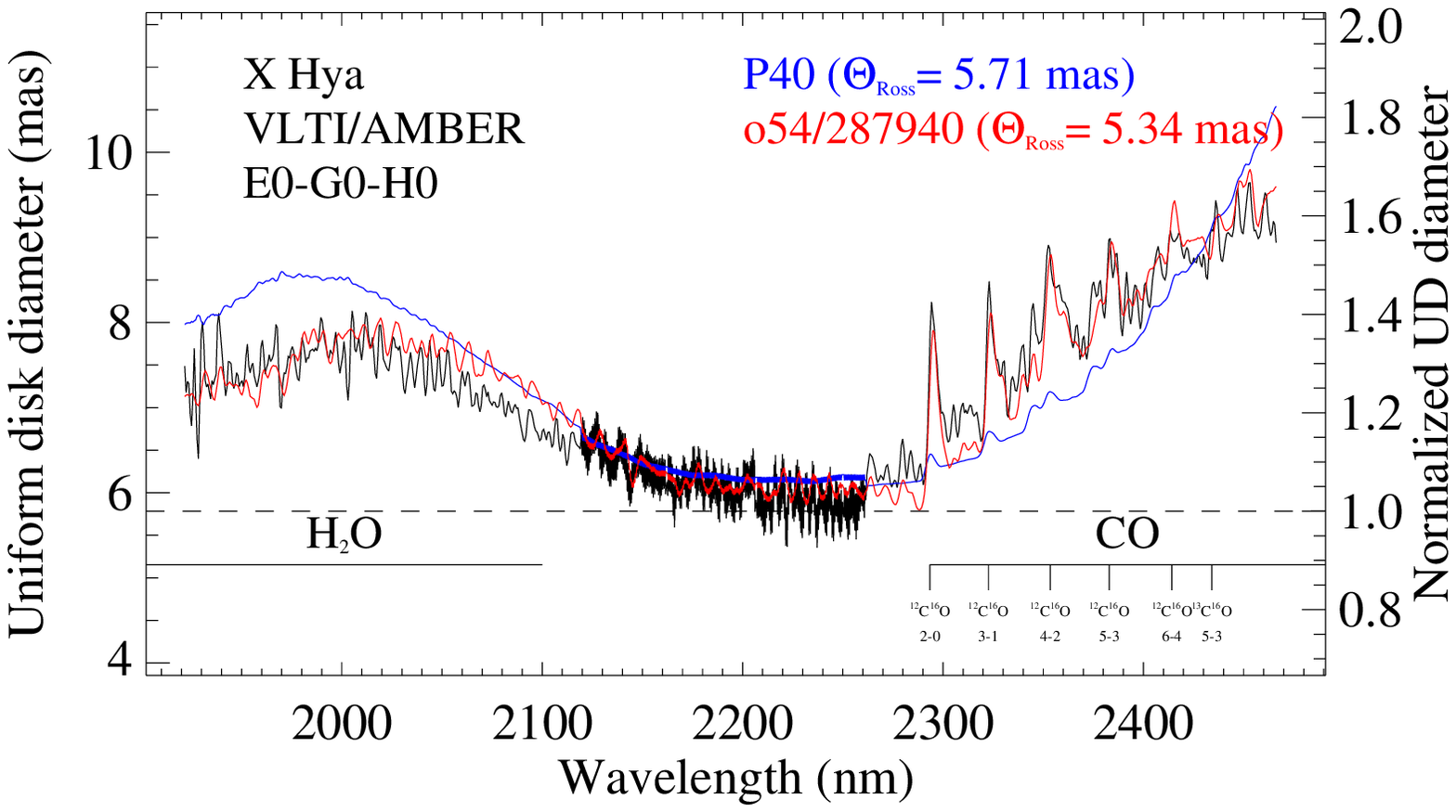}}
\resizebox{0.32\hsize}{!}{\includegraphics{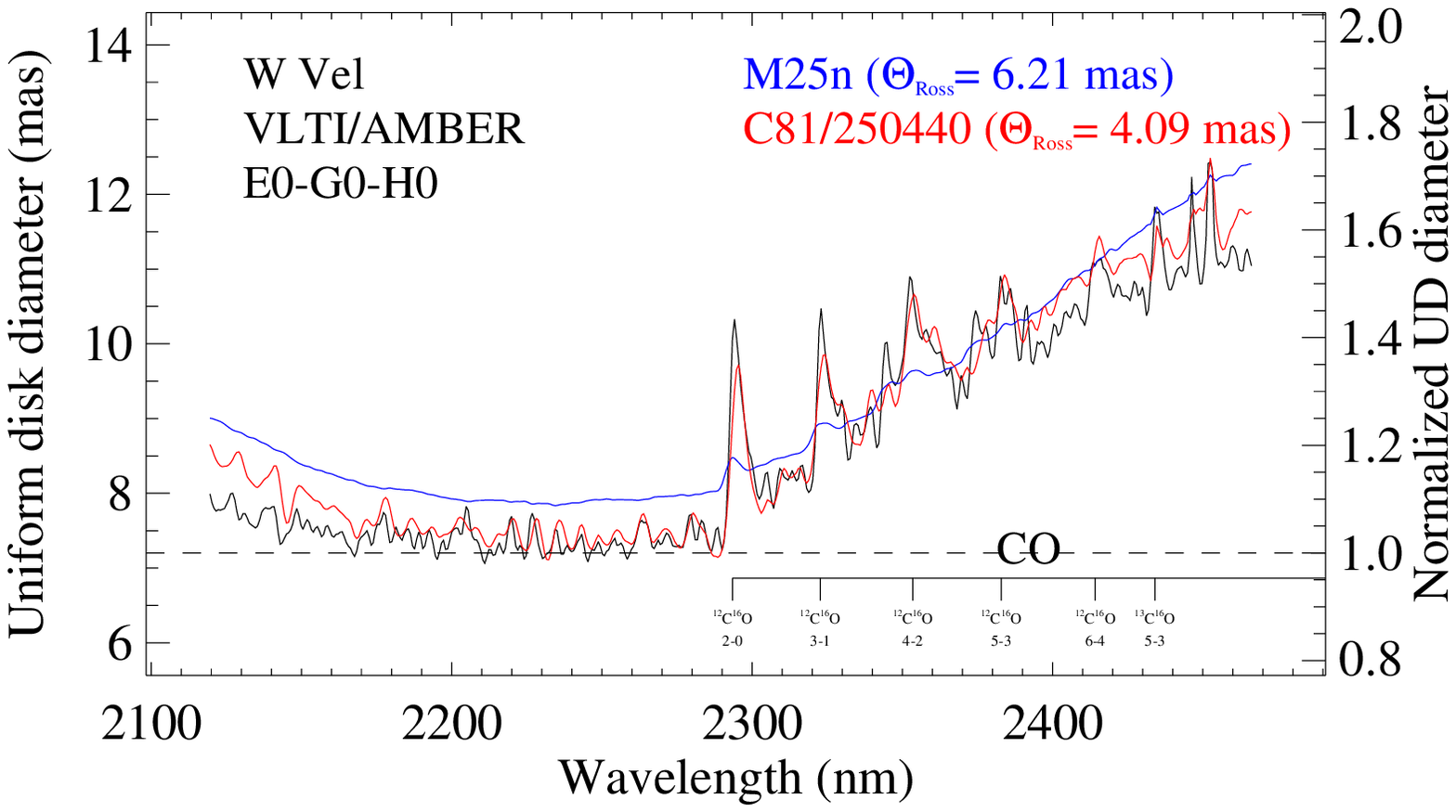}}
\resizebox{0.32\hsize}{!}{\includegraphics{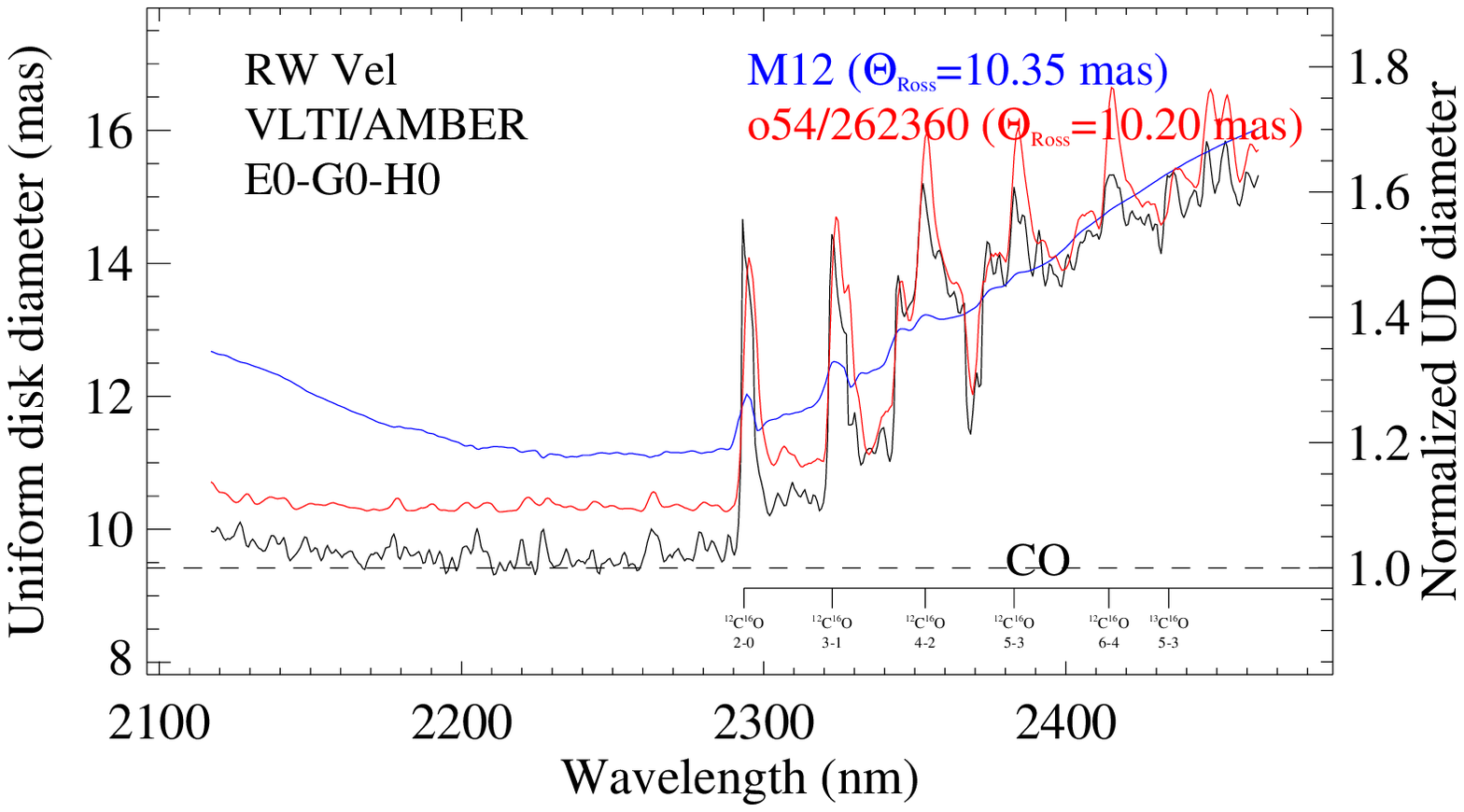}}

\caption{Flux, squared visibility amplitude, 
closure phase, and uniform disk diameter (from top to bottom)
as a function of wavelength for the Mira variables
X~Hya, W~Vel, and RW~Vel (from left to right). 
As Fig.~\protect\ref{fig:miras}, but for the remaining targets
of our sample.}
\label{fig:mirasonline}
\end{figure*}
}
Figures~\ref{fig:miras} \& \ref{fig:mirasonline} show the resulting AMBER flux, visibility, UD diameter,
and closure phase data. The UD diameter curve is obtained by fitting
a UD model to the visibility data separately for each spectral channel.
We note that the intensity profiles are expected to be 
more complex than a simple UD model, and this approach is used merely
to readily obtain a preliminary estimate of the
apparent diameter.
Signatures of a more complex model as well as asymmetries are 
most pronounced near and beyond the first visibility minimum, 
hence we used the two shorter baselines only for UD fits.

The AMBER data are compared to the {\tt P/M} dynamic model atmosphere series 
(Ireland et al. \cite{ireland04a,ireland04b}), as well as the newly
available {\tt CODEX} series (Ireland et al. \cite{ireland08,ireland11}). 
Both series are based on self-excited pulsation models.
The only fit parameter for each model atmosphere is the Rosseland angular diameter. 
Fig.~\ref{fig:miras} shows the result of the best-fit model of each of 
the {\tt P/M} and {\tt CODEX} series. Compared to the {\tt P/M} series, 
the {\tt CODEX} series uses the opacity sampling method, and the model series are 
available for a wider range of basic parameters of the underlying hypothetical 
non-pulsating parent star. The {\tt CODEX} intensity profile was tabulated in 
steps of 0.0005\,$\mu$m ($\lambda/\Delta\lambda=40000$ at $\lambda=2$\,$\mu$m),
and the synthetic visibility values averaged over the AMBER spectral 
channels assuming a Gaussian-shaped bandpass. The {\tt CODEX} series include 
the series {\tt o54} (parent star parameters: $M=1.1$\,M$\odot$, 
$L=5400$\,L$\odot$, $R=216$\,R$_\odot$, $P=330$ days), {\tt R52} 
(1.10\,M$\odot$, 5200\,L$\odot$, 209\,R$_\odot$, 307 d), {\tt C50} 
(1.35\,M$\odot$, 5050\,L$\odot$, 291\,R$_\odot$, 430 d), and {\tt C81} 
(1.35\,M$\odot$, 8160\,L$\odot$, 278\,R$_\odot$, 427 d), compared to the 
previous series {\tt P} (1.0\,M$\odot$, 3470\,L$\odot$, 241\,R$_\odot$, 332 d) 
and {\tt M} (1.2\,M$\odot$, 3470\,L$\odot$, 260\,R$_\odot$, 260 d).
Table~\ref{tab:results} provides an overview of the measured and
derived quantities of the four targets.

The targets of our sample have a visibility function with a characteristic
shape\footnote{The R~Cnc visibility reaches
into the second lobe, where the shape is inverted.}. 
For each baseline, it exhibits
a maximum near 2.25\,$\mu$m and decreases towards both shorter and
longer wavelengths. At shorter wavelengths, it shows a minimum 
near 2.0\,$\mu$m.
At longer wavelengths, between 2.3\,$\mu$m and 2.5\,$\mu$m, the visibility 
function displays sharp drops at the locations of the CO bandheads. This 
characteristic shape of the visibility function of the four targets studied 
here is consistent with low resolution ($\lambda/\Delta\Lambda\sim35$) 
AMBER data of the Mira variable S~Ori (Wittkowski et al. \cite{wittkowski08}), 
but shows additional details thanks to the increased spectral resolution.
Fig.~\ref{fig:miras} displays the position of the H$_2$O 
(1.9--2.1\,$\mu$m) and CO bands (2.3--2.5\,$\mu$m, but where H$_2$O may also 
contribute).

The corresponding wavelength-dependent uniform disk (UD) diameters show a 
minimum near the near-continuum bandpass at 2.25\,$\mu$m. They increase 
toward 2.0\,$\mu$m, where the H$_2$O opacity is large, by up to $\sim$\,30\%. 
The CO bandheads between 2.29\,$\mu$m and 2.48\,$\mu$m are clearly visible for 
all targets with an increase in the UD diameter by up to $\sim$\,70\%.

The best-fit models of the {\tt P/M} and {\tt CODEX} model series
predict visibility and UD curves that are consistent 
with the observations. Here, the newly available models of the {\tt CODEX} 
series provide a closer agreement with the data than the {\tt P/M} series, 
at both the locations of the CO bandheads, which may be expected because of 
the newly introduced opacity sampling method and their higher spectral 
resolution, and in terms of the overall shape of the curve. The latter 
may be explained by the availability of more model series with additional 
parent star parameters and with an increased phase coverage per series. 
The AMBER flux curves are consistent with those of the dynamic model 
atmosphere series. In addition, the pulsation phases 
of the best-fit {\tt CODEX} models are consistent with the observed visual 
phases, and their effective temperature is consistent with that derived from 
the fitted angular Rosseland-mean diameter and the bolometric flux. 
Furthermore, the distances obtained from the fitted angular diameter and the 
model radius are consistent with those derived from the measured apparent
bolometric magnitude and the model luminosity, as well as with the 
period-luminosity distance from Whitelock et al. (\cite{whitelock08},
cf. Table~\ref{tab:results}) within 1--3\,$\sigma$.

The closure phase functions of our targets exhibit significant 
wavelength-dependent non-zero values at all wavelengths. Non-zero
values of the closure phase are indicative of
deviations from point symmetry, here along the projected 
intensity profile onto the orientation of the position angle given in 
Table~\ref{tab:vlti}. The consistency of the visibility amplitudes with 
spherical models and the closure phase deviation being more 
significant for targets that are well-resolved indicates that the deviation 
from point symmetry originates from sub-structure at a relatively low
flux level of an overall spherical
intensity distribution. The strongest closure phase signal is obtained for 
the most clearly resolved target R Cnc. It 
differs strongly from point-symmetry in the H$_2$O band at 2.0\,$\mu$m  
with a value of 110\degr\ $\pm$ 4\degr. At the continuum bandpass at 
2.25\,$\mu$m, the closure phase value is closer to 0 with a value of 
45\degr\ $\pm$ 5\degr. Towards longer wavelengths (CO band), the average value 
is 153\degr\ $\pm$ 4\degr, with values up to 165\degr\ in the bandheads, 
and thus close to a 180\degr\ symmetric value. As an illustration, the R~Cnc closure 
phase value in the H$_2$O band can be caused by the addition of an unresolved 
(up to $\sim$ 3 mas diameter) spot to a circular disk, contributing up to $\sim$ 3\% of 
the total flux, i.e. corresponding to the level of the visibility departures between
data and spherical models, at a separation of $\sim$ 4\,mas.
\begin{table*}
\caption{Results}
\label{tab:results}
\centering
\begin{tabular}{lrrrr}
\hline\hline
                                                  & R Cnc                  &    XHya             &      W Vel           &  RW Vel              \\\hline
$\phi_\mathrm{Vis}$ ($P$)                         & 0.3 (356 d)            & 0.7 (298 d)         & 0.9 (397 d)          & (452 d)              \\[1ex]
UD, full $\lambda$      [mas]      ($\chi^2_\nu$) & 15.2 $\pm$ 0.1 (17.2)  & 7.8 $\pm$ 0.1 (1.7) & 10.3 $\pm$ 0.1 (4.0) & 14.0 $\pm$ 0.1 (6.3) \\
UD, 2.25 $\mu$m (cont.) [mas]      ($\chi^2_\nu$) & 13.2 $\pm$ 0.1 (1.3)   & 6.2 $\pm$ 0.1 (0.3) & 8.0 $\pm$ 0.1  (0.2) & 10.6 $\pm$ 0.1 (0.3) \\
UD, 2.0 $\mu$m (H$_2$O) [mas]      ($\chi^2_\nu$) & 14.7 $\pm$ 0.1 (4.8)   & 8.3 $\pm$ 0.1 (0.4) & not obtained         & not obtained         \\
UD, 2.4 $\mu$m (CO)     [mas]      ($\chi^2_\nu$) & 16.1 $\pm$ 0.1 (8.6)   & 9.0 $\pm$ 0.1 (0.5) & 11.2 $\pm$ 0.1 (1.0) & 14.6 $\pm$ 0.1 (1.9) \\[1ex]
CP, full $\lambda$      [deg]                     & 110 $\pm$ 4            &  5 $\pm$ 2          & -8 $\pm$ 2           & 14 $\pm$ 3           \\
CP, 2.25 $\mu$m (cont.) [deg]                     & 45 $\pm$ 5             &  2 $\pm$ 2          & -4 $\pm$ 2           &  3 $\pm$ 2           \\
CP, 2.0 $\mu$m (H$_2$O) [deg]                     & 108 $\pm$ 4            & 11 $\pm$ 3          & not obtained         & not obtained         \\
CP, 2.4 $\mu$m (CO)     [deg]                     & 153 $\pm$ 4            &  5 $\pm$ 3          & -12 $\pm$ 3          &  34 $\pm$ 4          \\[1ex]
Best {\tt P/M} Model ($\chi^2_\nu$)    & M11n (7.4) &P40 (1.9)& M25n (4.8) & M12 (9.5) \\
Best {\tt CODEX} Model ($\chi^2_\nu$)  & C50-378680 (4.6) &o54-287940 (0.7)& C81-250440 (1.6)&  o54-262360 (4.6)\\
$\Theta_\mathrm{Ross}$  [mas]                     & 11.8 $\pm$ 0.7         & 4.9 $\pm$ 0.8       & 5.8 $\pm$ 1.2        & 9.2 $\pm$ 1.3        \\
$T_\mathrm{eff}^\mathrm{Model}$ [K]               & 2628 $\pm$ 184         & 3064 $\pm$ 380      & 3216 $\pm$ 225       & 3142 $\pm$ 243       \\
$\Phi_\mathrm{Model}$                             & 0.45 $\pm$ 0.09        & 0.67 $\pm$ 0.28     & 0.77 $\pm$ 0.12      & 0.73 $\pm$ 0.10      \\[1ex]
$T_\mathrm{eff}^\mathrm{Meas}$  [K]               & 2694 $\pm$ 140         & 2912 $\pm$ 310      & 2990 $\pm$ 390       & $m_\mathrm{bol}$ not avail. \\[1ex]
 $d_\mathrm{PL}$         [pc]                      & 280 $\pm$ 28           & 440 $\pm$ 44        & 510 $\pm 51$         & 366 $\pm$ 36          \\
 $d_\mathrm{Model}$ ($\Theta_\mathrm{Ross}$, $R_\mathrm{Ross}^\mathrm{Model}$) [pc]  & 252 $\pm$ 28 & 465 $\pm$ 49 & 362 $\pm$ 96 & 217 $\pm$ 33 \\
 $d_\mathrm{Model}$ ($m_\mathrm{bol}$, $M_\mathrm{bol}^\mathrm{Model}$)        [pc]  & 246 $\pm$ 46 & 503 $\pm$ 80 & 400 $\pm$ 37 & $m_\mathrm{bol}$ not avail.   \\\hline
\end{tabular}
\tablefoot{The visual phase $\Phi_\mathrm{Vis}$ and period $P$ was
obtained from fitting a sine function to the AAVSO and AFOEV data
of the last ten periods. For RW Vel such data are not available,
and we list the GCVS (Samus et al. \cite{samus09}) period instead.
The UD diameters are obtained from a fit to all data
(full $\lambda$), to the continuum bandpass 2.24--2.26\,$\mu$m
(2.25\,$\mu$m), to the H$_2$O bandpass 1.95--2.05\,$\mu$m (2.0\,$\mu$m),
and to the CO bandpass 2.35--2.45\,$\mu$m (2.4\,$\mu$m). The 
closure phases CP were averaged over the same bandpasses. The Rosseland mean
angular diameter $\Theta_\mathrm{Ross}$ is an average of the fitted diameters
based on several closely fitting CODEX model atmospheres of any series with
average effective temperature $T_\mathrm{eff}^\mathrm{Model}$ and
an visual phase $\Phi_\mathrm{Model}$.
$T_\mathrm{eff}^\mathrm{Meas}$ was derived from 
$\Theta_\mathrm{Ross}$ and $m_\mathrm{bol}$ (Tab. \protect\ref{tab:saao}).
$d_\mathrm{PL}$ is the period-luminosity distance from Whitelock 
et al. (\cite{whitelock08}). RW~Vel is based on the same scale,
but using the GCVS period and the mean $K$ magnitude from Smith et al.
(\cite{smith02}).
The last two rows give the distances based on $\Theta_\mathrm{Ross}$
and $R_\mathrm{Ross}^\mathrm{Model}$ of the models at the respective
phase, and based on $m_\mathrm{bol}$ and $M_\mathrm{bol}^\mathrm{Model}$.
}
\end{table*}
\section{Discussion}
The four Mira variables of our sample exhibit consistent characteristic 
wavelength dependences of the visibility and consequently the corresponding uniform 
disk diameter that are consistent with those of earlier low-resolution AMBER data of 
the Mira variable S~Ori and the predictions of the {\tt P/M} and {\tt CODEX} 
dynamic model atmosphere series. Here, the newly available {\tt CODEX} series 
provides a closer agreement with the data than the earlier {\tt P/M} series. 
This result confirms that the wavelength-dependent angular 
diameter is caused by the atmospheric molecular layers, here most importantly 
H$_2$O and CO, as they are naturally included in the dynamic model atmosphere 
series. Concurrent $JHKL$ photometry obtained at the SAAO was used to derive
$T_\mathrm{eff}$ based on the integrated bolometric flux and the fitted 
Rosseland-mean angular diameter. Parameters of the best-fit model 
atmospheres, such as visual phase, effective temperature, and distances are 
consistent with independent estimates, which provides additional 
confidence in the {\tt CODEX} modeling approach.

The closure phase functions of our targets exhibit non-zero values at all 
wavelengths with a wavelength dependence that also correlates with the 
positions of the H$_2$O and CO bands. This result indicates a complex 
non-spherical stratification of the extended atmosphere of Mira variables.
The most significant deviation from point symmetry is observed in the 
H$_2$O band around 2.0\,$\mu$m, in particular for the clearly resolved 
target R~Cnc. We interpret this signal as an indication of inhomogeneities 
or clumps within the water vapor layer on scales of a few resolution elements across 
the stellar disk. These inhomogeneous water shells have also been detected for 
the Mira variable U Ori (Pluzhnik et al. \cite{pluzhnik09}) and 
the symbiotic Mira variable R Aqr (Ragland et al. \cite{ragland08}).
The deviations from point symmetry at the near-continuum bandpass may be related
either to photospheric convection cells (cf. Freytag \& H\"ofner \cite{freytag08}) 
or the inhomogeneities of 
molecular layers that may also contaminate this bandpass. 

Inhomogeneous or clumpy molecular layers may be expected as the
result of chaotic motion induced by the interaction of pulsation and shock
fronts with the extended atmosphere. The CODEX models indicate that
the outer mass zones outward of 1.5--2 Rosseland radii are only loosely
connected to the stellar pulsation (cf. Fig. 1 of Ireland et al. 
\cite{ireland08}). Likewise, outer mass zones on different sides of the star
may be only weakly correlated with each other and may have different 
extensions. Icke et al. (\cite{icke92}) described as well that the 
outer layers of an evolved AGB star may respond with chaotic motion
to the pulsations that originate in the stellar interior.
Observations of water vapor layers may be particularly sensitive to these 
effects, but other molecules, such as CO, that are expected to be plentiful in the 
shocked region of the atmosphere (cf. Cherchneff \cite{cherchneff06}) would 
also be affected by this large-scale chaotic motion.
The resulting clumpy structure of molecular layers may explain our complex 
wavelength-dependent closure phase signal.

Further interferometric campaigns with high spatial and spectral
resolution are clearly needed to characterize in detail the morphology of 
atmospheric molecular layers in Mira variables.
\begin{acknowledgements}
This research has made use of the  \texttt{AMBER data reduction package} of the
Jean-Marie Mariotti Center.
We acknowledge with thanks the variable star observations from the AAVSO 
International Database contributed by observers worldwide and used in 
this research.
This research has made use of the AFOEV database, operated at CDS, France.
\end{acknowledgements}
\end{document}